\documentclass{article}
\usepackage{glas}
\usepackage{epsf}
\epsfverbosetrue
\def\Journal#1#2#3#4{{#1} {\bf #2}, (#4) #3}
\def\PRD{{\em Phys. Rev.} D}

\def\PLB{{\em Phys. Lett.}  B}
\def\NPB{{\em Nucl. Phys.} B}

\def\delz{\mbox{$<\!\!\delta Z\!\!>$}} 

\newcommand{\eT}{\mbox{$E_T$}}

\newcommand{\eTg}{\mbox{$E_T^{\,\gamma}$}}

\begin{document}
\begin{titlepage}{GLAS-PPE/2000--07}{June 2000}
\title{Prompt Photons in Photoproduction at HERA}
\author{Sung--Won Lee}
\vspace{-0.5cm}
\collaboration{for the ZEUS Collaboration}
\begin{abstract}
First inclusive measurements of isolated prompt photons in
photoproduction at HERA have been made with the
ZEUS detector. Cross sections are given as a function of the
pseudorapidity and the transverse energy of
the photon, for $\eTg > $ 5 GeV in the $\gamma p$ centre-of-mass
energy range 134--285 GeV. Comparisons are made with predictions
from LO Monte Carlo models and NLO QCD calculations.  
For forward $\eta^\gamma$ (proton direction) good agreement 
is found, but in the rear direction all predictions fall below 
the data.
\end{abstract}
%\vfill
\conference{Talk given at the 8th International Workshop \\on Deep 
Inelastic Scattering and QCD (DIS '2000), \\
Liverpool, UK, April 25--30, 2000}
\end{titlepage}

\section{Introduction}

Isolated high transverse energy (``prompt'') photon processes at HERA 
could yield information about the quark and gluon content of the photon,
together with the gluon structure of the proton.
The particular virtue of prompt photon processes is that 
the observed final state photon emerges directly from a QCD diagram 
without the subsequent hadronisation which complicates the study of 
high $E_T$ quarks and gluons. 

In a ZEUS paper~\cite{prph_ref1}
the observation of prompt photons was first confirmed at HERA.
More recently~\cite{prph_ref2}, ZEUS collaboration has measured
the cross sections of inclusive prompt photons in photoproduction 
reactions, using an integrated luminosity of 38.4 pb$^{-1}$. 
Comparisons are made with predictions from Monte Carlo models containing 
leading-logarithm parton showers, and with next-to-leading-order QCD 
calculations, using currently available parameterisations of the photon
structure.

\section{Evaluation of the photon signal}

The data used here were obtained from  $e^+ p$ running in 1996--97 at
HERA, with $E_e = 27.5$ GeV, $E_p = 820$ GeV. 

The major components in the analysis are the 
central tracking detector(CTD) and the uranium calorimeter(UCAL).
Prompt photons are detected in the barrel section of the calorimeter, 
which consists of an electromagnetic section (BEMC) followed by two 
hadronic sections. It enable a partial discrimination between single 
$\gamma$ signals and the decay product of neutral mesons. 
A typical high-\eT\ photon signal is observed in a small cluster of
BEMC cells, with no associated CTD track. An isolation cone was also 
imposed around photon candidates within a cone of unit radius in 
$(\eta,\,\phi)$, to reduce backgrounds from dijet events with part
of one jet misidentified as a single photon.

\begin{figure}[htb]
\vspace{-1cm}
\centerline{
\epsfig{figure=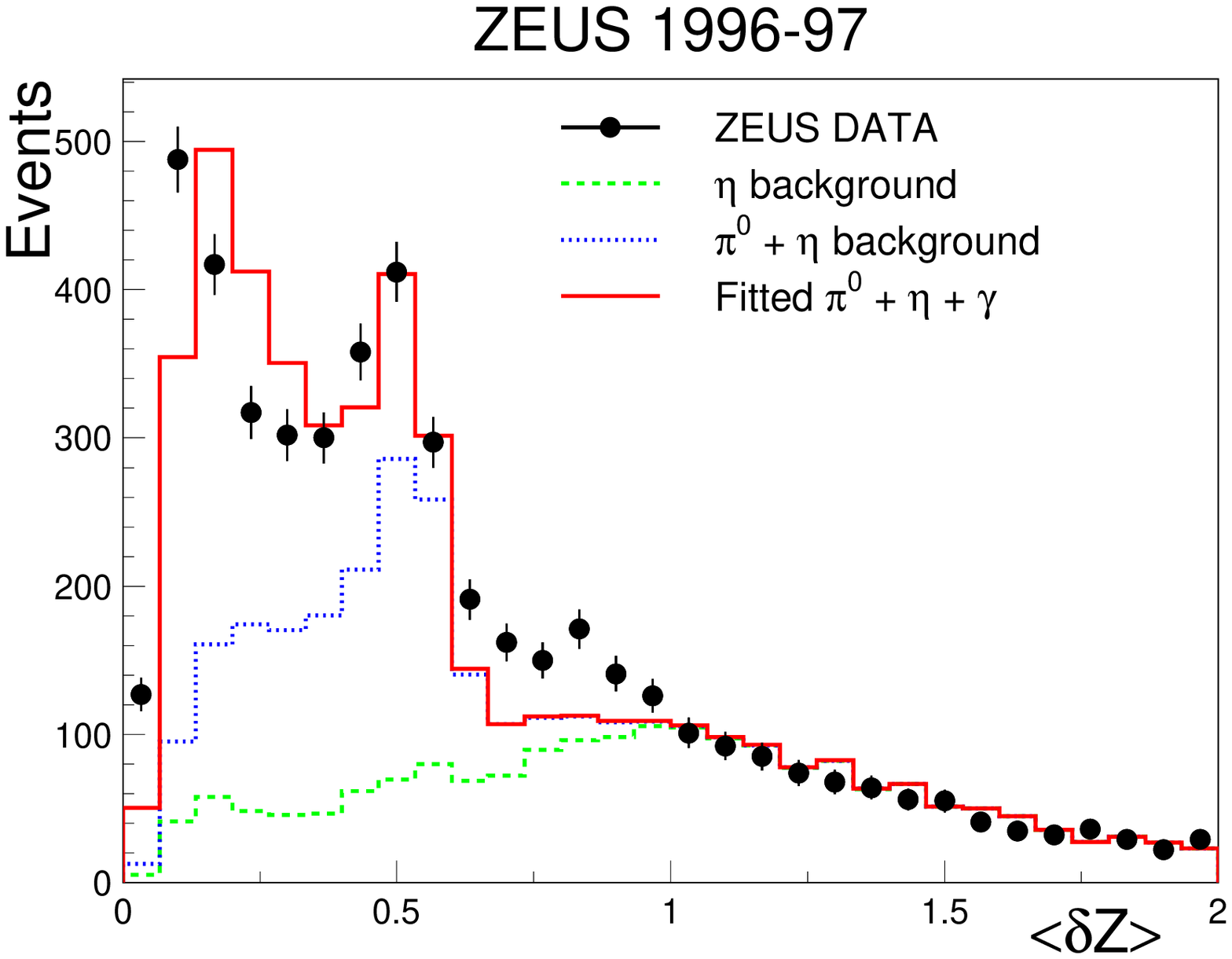,height=3.5in}
\hspace{-0.3cm}
\epsfig{figure=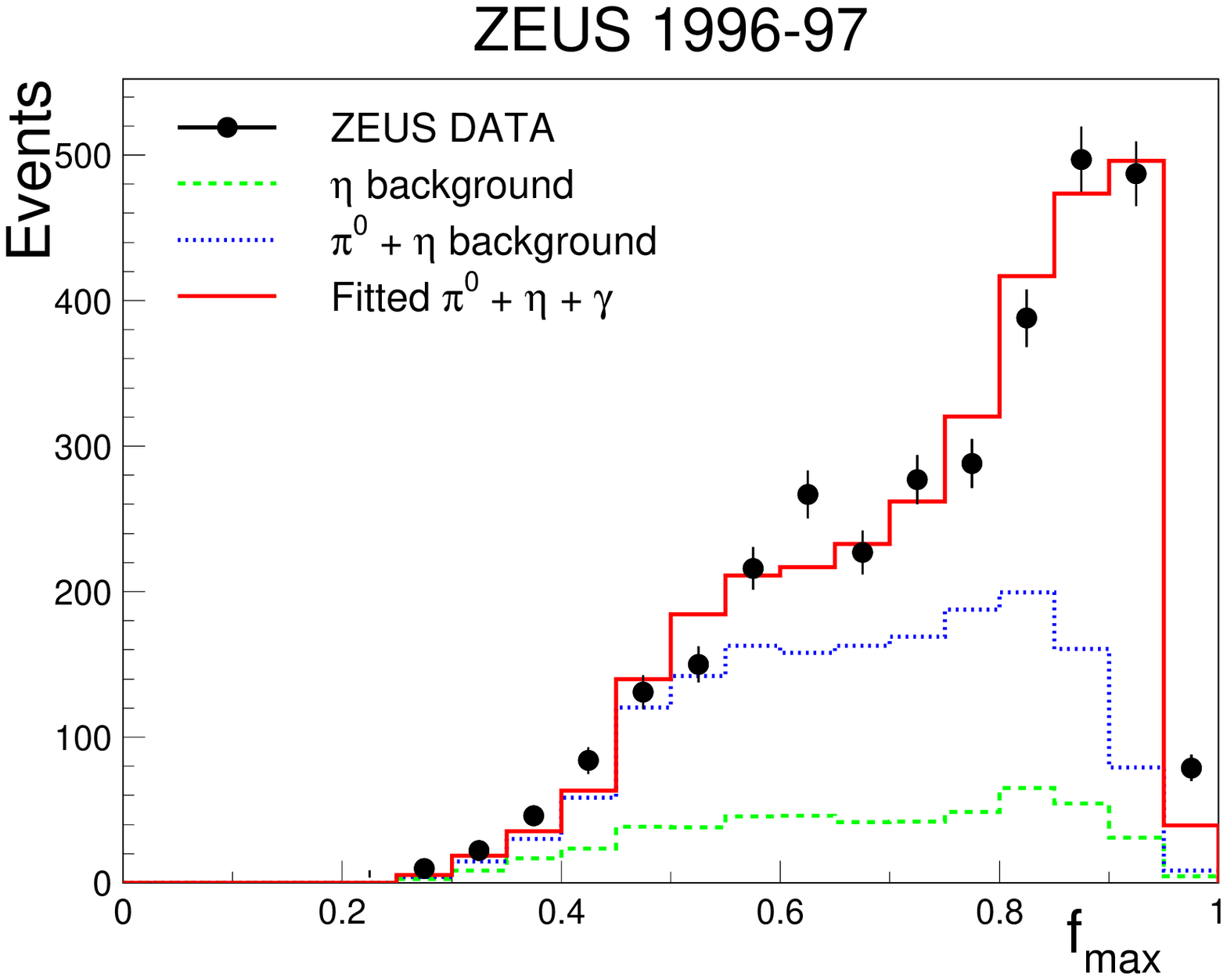,height=3.5in}}
\vskip -1cm
\caption{Distribution of (a) \delz\ and (b) $f_{max}$ for prompt photon
candidates in selected events. Also given in both cases are fitted MC 
distributions for $\gamma$, $\pi^0$ and $\eta$ mesons.\label{fig:shower}}
\end{figure}

Two shape-dependent quantities were studied in order to further 
distinguish $\gamma$, $\pi^0$ and $\eta$ signals.  
These were (1) the mean width $<\!\!\delta Z\!\!>$\ of the BEMC 
cluster in $Z$ and (2) the fraction $f_{max}$ of the cluster energy 
found in the most energetic cell in the cluster.
The \delz\ distribution is shown in Figure 1(a),
in which peaks due to the $\gamma$ and $\pi^0$ contributions are 
clearly visible. The tail quantified the $\eta$ background; photon 
candidates in this region were removed.

The extraction of the photon signal from the mixture of photons 
and a neutral meson background was done by means of the $f_{max}$
distribution. Figure 1(b) shows the shape of the $f_{max}$ distribution 
for the final event sample, after the \delz\ cut, 
fitted to the  $\eta$ component determined from the \delz\
distribution and freely-varying $\gamma$ and $\pi^0$ contributions. 
Above an $f_{max}$ value of 0.75, the distribution is dominated by the
photons; below this value it consists mainly of meson background.
The numbers of candidates with $f_{max} \ge 0.75$ and $f_{max} < 0.75$
were calculated for the sample of events occurring in each bin of any
measured quantity. From these numbers, and the ratios of the
corresponding numbers for the $f_{max}$ distributions of the
single particle samples, the number of photon events in the given bin
was evaluated.  Further details of the background subtraction method
are given in reference.~\cite{prph_ref1,prph_ref2}

\begin{figure}
\centerline{
\epsfig{figure=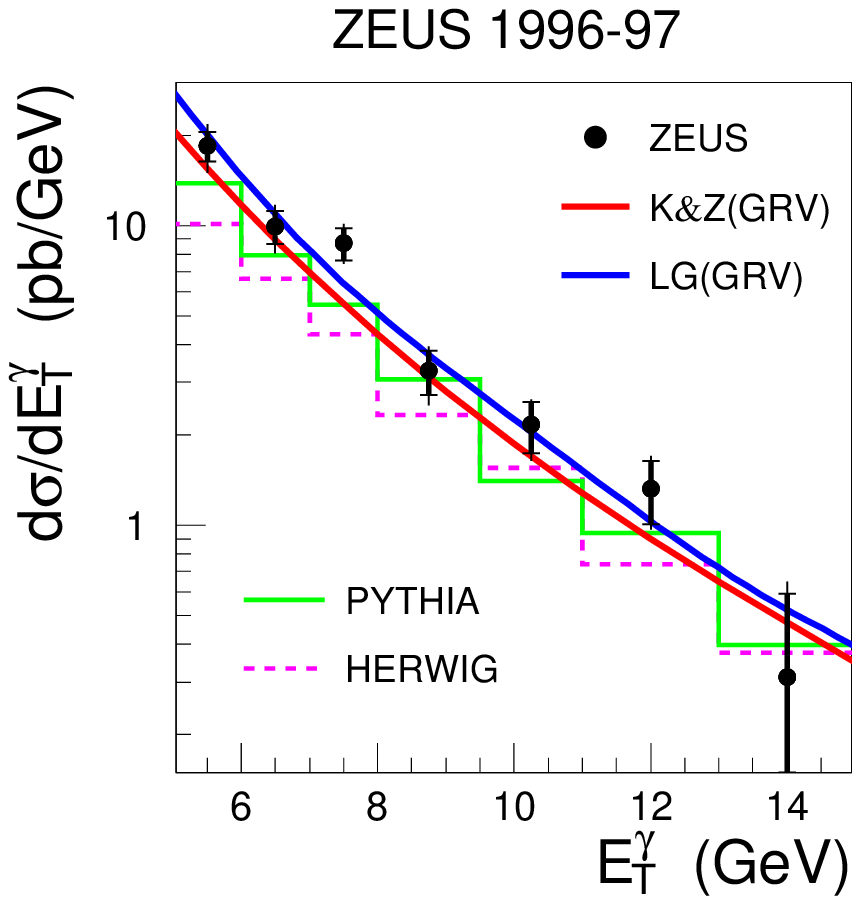,height=3.0in}
\hspace{0.15cm}
\epsfig{figure=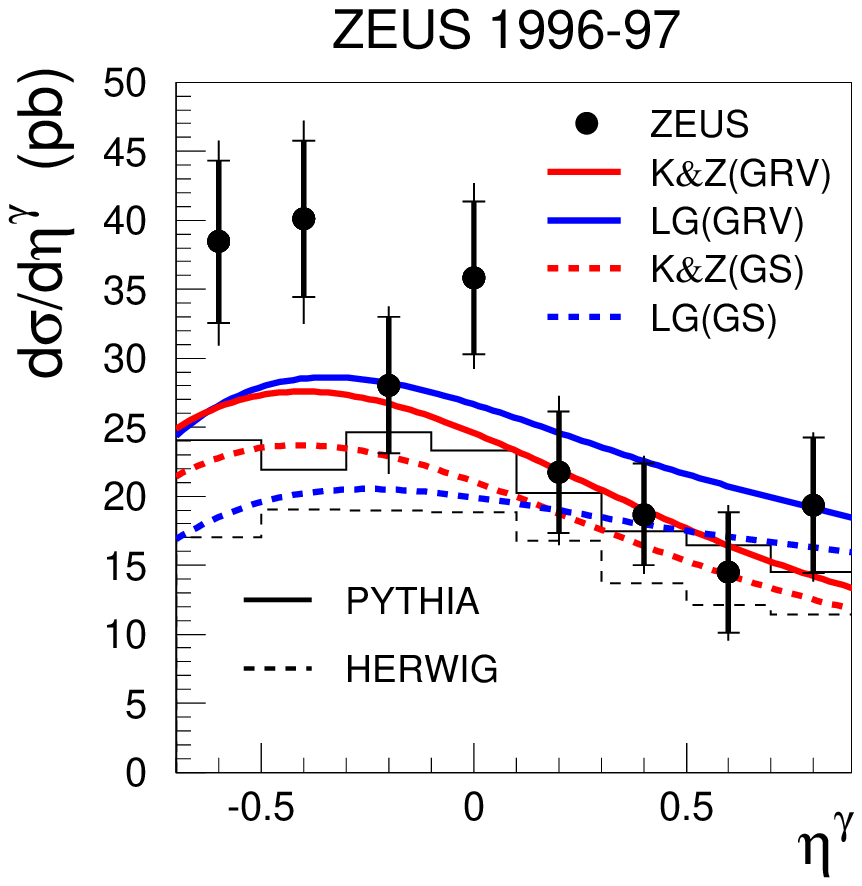,height=3.0in}}
\caption{Differential cross section (a) $d\sigma/d\eTg$ 
and (b) $d\sigma/d\eta^\gamma$ for isolated prompt photons
produced over $-0.7 < \eta^\gamma < 0.9$. \label{fig:xsec}}
\end{figure}

\section{Results}

We evaluate cross sections for prompt photon production corrected by
means of PYTHIA using GRV photon structure functions~\cite{grv}.
A bin-by-bin correction factors were applied to the detector--level
measurements so as to correct to cross sections in the $\gamma p$ 
centre--of--mass energy 134 -- 285 GeV. 
The systematic error of 15\% were taken into account and
were finally combined in quadrature. The main contributions are
from the energy scale of the calorimeter and the background subtraction.
In presenting cross sections, comparison is made with two types of
theoretical calculation, in which the pdf sets taken for both the
photon and proton can be varied. These are (1) PYTHIA and HERWIG 
calculations evaluated at the final-state hadron level and
(2) NLO parton--level calculations of Gordon~\cite{gordon}(LG)
and of Krawczyk and Zembrzuski~\cite{warsaw}(K\&Z).

 Figure 2 (a) gives the inclusive cross--section
$d\sigma/d\eTg$ for isolated prompt photons in the range
$-0.7 < \eta^\gamma < 0.9$. All the theoretical models describe 
the shape of the data well; however the predictions of PYTHIA 
and especially HERWIG are too low in magnitude.
The LG and K\&Z calculations give better agreement with the data.

 The inclusive cross--section $d\sigma/d\eta^\gamma$ for
isolated prompt photons in the range $5 <\eTg < 10$ GeV is
shown in figure 2 (b) and compared with theoretical calculations,
using two sets of photon pdf, GS~\cite{gs}and GRV.
The LG and K\&Z calculations gives a good description of the data
for forward (proton direction) $\eta^\gamma$ range and are similar
to PYTHIA prediction. However all the calculations lie below the 
data in the lower $\eta^\gamma$ range, where the curves using the
GS parton densities give poorer agreement than those using GRV.

 The discrepancy between data and theory at negative $\eta^{\gamma}$
is found to be relatively strongest at low $\gamma p$ centre-of-mass
energy range. In the lowest $W$ range (134--170~GeV), both theory 
and data show a peaking at negative $\eta^\gamma$, but it is stronger
in the data. In the highest $W$ range (212--285~GeV), 
agreement is found between theory and data.  The movement of the
peak can be qualitatively understood by noting that for fixed values
of \eT\ and $x_\gamma$, where $x_\gamma$ is the fraction of the
incident photon energy that contributes to the resolved QCD
subprocesses, measurements at increasing $y$ correspond on average to
decreasing values of pseudorapidity.  By varying the theoretical
parameters, the discrepancy was found to correspond in the K\&Z
calculation to insufficient high $x_\gamma$ partons in the resolved
photon.

\section{Conclusions}

 The photoproduction of isolated prompt photons within the $\gamma p$
centre-of-mass energy range 134--285 GeV has been measured in the  
ZEUS detector at HERA. Inclusive cross sections have been presented 
as a function of \eTg\ for photons in $-0.7 < \eta^\gamma < 0.9$, 
and as a function of $\eta^\gamma$ for photons with $5 < \eTg < 10$ GeV. 

 Comparisons have been made with predictions from LO Monte Carlos,
and from NLO calculations. The models are able to describe the data
well for forward $\eta^\gamma$, but are low in the rear direction. 
None of the available variations of the model parameters was found 
to be capable of removing the discrepancy with the data. This result 
would appear to indicate a need to review the present theoretical 
modelling of the parton structure of the photon.

\section*{Acknowledgments}
We are grateful to L. E. Gordon, Maria Krawczyk and Andrzej Zembrzuski
for helpful conversations, and for providing theoretical calculations.


\begin{thebibliography}{99}

\bibitem{prph_ref1} 
ZEUS Colla., J. Breitweg et al., \Journal{\PLB}{413}{201}{1997}.

\bibitem{prph_ref2} 
ZEUS Colla., J. Breitweg et al., \Journal{\PLB}{472}{175}{2000}. 

\bibitem{grv} 
M. Gl\"uck, E. Reya and A. Vogt, \Journal{\PRD}{46}{1973}{1992}.

\bibitem{gs} 
L. E. Gordon and J. K. Storrow, \Journal{\NPB}{489}{405}{1997}.

\bibitem{gordon} 
L. E. Gordon, \Journal{\PRD}{57}{235}{1998}.

\bibitem{warsaw} 
M. Krawczyk and A. Zembrzuski, {\tt hep-ph/9810253}.

\end{thebibliography}
\end{document}